# Investigation of a Bit-Sequence Reconciliation Protocol Based on Neural TPM Networks in Secure Quantum Communications

Matvey Yorkhov
*Department of Complex Security of Computing Systems*
Tomsk State University of Control Systems and Radioelectronics
Tomsk, Russia
matveyorkhov@gmail.com

Vladimir Faerman
*Department of Complex Security of Computing SystemsTomsk State University of Control Systems and Radioelectronics*
Tomsk, Russia
fva@fb.tusur.ru

Anton Konev
*Department of Complex Security of Computing Systems Tomsk State University of Control Systems and Radioelectronics*
Tomsk, Russia

*Abstract*— The article discusses a key reconciliation protocol for quantum key distribution (QKD) systems based on Tree Parity Machines (TPM). The idea of transforming key material into neural network weights is presented. Two experiments were conducted to study how the number of synchronization iterations and the amount of leaked information depend on the quantum bit error rate (QBER) and the range of neural network weights. The results show a direct relationship between the average number of synchronization iterations and QBER, an increase in iterations when the weight range is expanded, and a reduction in leaked information as the weight range increases. Based on these results, conclusions are drawn regarding the applicability of the protocol and the prospects for further research on neural cryptographic methods in the context of key reconciliation.

## I. Introduction

Cryptography is an important tool for ensuring information security, allowing the prevention of unauthorized access and the protection of data confidentiality.

Quantum key distribution (QKD) is a relatively new and rapidly developing field, whose importance is increasing with the advancement of quantum computing [1]. A crucial aspect of QKD is the post-processing of the key sequence distributed over the quantum channel, which involves correcting randomly occurring errors on both sides of the exchange.

Historically, the first proposed error-correction algorithm was the BB84 protocol [2], which served as a model for many well-known protocols studied in the literature and used today [3,4]. However, there are protocols based on other classes of computational methods that have not gained wide adoption and have not been systematically studied. The present work focuses on the study of a key reconciliation protocol based on a Tree Parity Machine (TPM) network, applied in cryptography.

## II. Reconcilation Method

### A. Tree Parity Machine as a Premitive

Neural cryptography is based on the use of neural networks for encryption tasks. The most common architecture applied in cryptography is the Tree Parity Machine (TPM) [5].

This network consists of input, hidden, and output layers. The input layer contains $K \times N$ neurons, where $K$ is the number of neurons in the hidden layer and $N$ is the number of inputs for each hidden-layer neuron (Fig. 1) [6]. The output layer consists of a single neuron.

The method of key reconciliation using TPM was proposed in [7] and is described below.

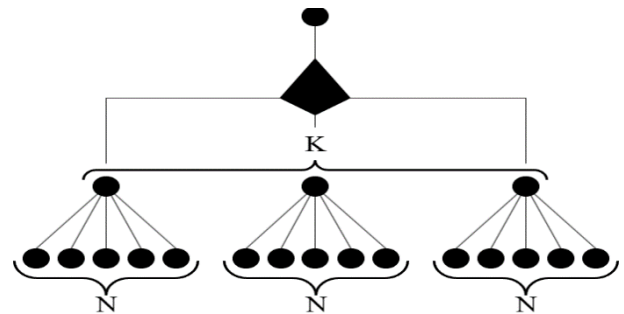

Fig. 1. Scheme of the TPM architecture.

### B. Algorithmic Description

The TPM-based key reconciliation algorithm enables two parties to correct errors in distributed key sequences using neural networks. It relies on synchronizing Tree Parity Machine networks on both sides, transforming bit strings into network weights, and iteratively adjusting them until the networks become identical. Once synchronization is complete, the weights are converted back into matching bit sequences, effectively reconciling the keys while minimizing information leakage.

First, the sender (Alice) and the receiver (Bob) construct TPM networks based on their bit strings, converting the bits into weights within the range $[-L; L]$ which define the connections between the input and hidden layers. The parameters $L$, $N$, and $K$ are agreed upon in advance. As a result, networks with the same structure and similar sets of weight coefficients are created, with differences arising only from bit errors in key distribution (QBER).

Second, after network initialization, synchronization is performed. In each synchronization round, Alice generates a random input vector of the required length and sends it to Bob along with the output bit. Bob calculates the output bit on his side and compares it with the bit received from Alice. If the output bits do not match, Alice performs a new iteration within the same round by generating a new input vector. If the bits match, training is performed using Hebbian, anti-Hebbian, or random-walk algorithms [7]. After adjusting the weights on both sides, the current round ends. Synchronization rounds are repeated until the weights become identical.

Finally, when the weights of the TPM networks on both sides match, Alice and Bob perform the reverse transformation of the weights back into a bit sequence. The resulting strings will be identical if the networks were successfully synchronized [7].

A block diagram for this method is shown in Fig. 2.

*C. Mapping of the Input Sequence into Weights*

An important aspect of the practical implementation of the algorithm is the introduction of a reversible operation that transforms a reconciled bit string of a given length into a set of weight coefficients. One possible solution is an algorithm in which the key sequence is divided into blocks of $b$ bits each. Each block is then converted into a decimal number, from which half of the maximum value for the given $b$ is subtracted. So the half-width of the weight range $L$ is calculated as

$$L = 2^{b-1}. \qquad (1)$$

As a result, the reconciled bit string is mapped to weight coefficients, each belonging to the range $[-L, \ldots, L-1]$.

A drawback of this approach is the asymmetry of the range, which is expressed in the unequal number of negative and positive values. This does not affect network synchronization, as it does not impact the ability to adjust the weight coefficients. However, according to [8], it influences the probability distribution of coefficients obtained after adjustment. This reduces the entropy of the sequence resulting from the reverse transformation of the weights back into a bit string. Nevertheless, the effect is minor for sufficiently large $L$, and it does not outweigh the practical advantages provided by the simplicity of the transformation [8].

*D. Security Considerations*

The security of the protocol was analyzed in [9]. Formal analysis methods showed that it is resistant to man-in-the-middle, pendulum, and summation-synchronization attacks. From a security perspective, the protocol appears promising.

In TPM-based key reconciliation, synchronizing neural network weights provides an additional layer of security. Because the weights are adjusted iteratively and the networks themselves are never directly exposed, an eavesdropper cannot easily reconstruct the key from intercepted synchronization data. Any attempt to interfere with the synchronization process introduces detectable discrepancies, making the protocol inherently robust against passive attacks.

Furthermore, the probabilistic nature of neural network training and the randomness of input vectors make active attacks, such as attempting to force synchronization or predict weight updates, extremely difficult. Even if an attacker knows the algorithm and network parameters, the uncertainty introduced by bit errors in the initial key sequence and the random weight updates ensures that the final key remains confidential. These properties highlight the potential of TPM-based methods as a secure alternative to classical error-correction protocols in quantum key distribution systems.

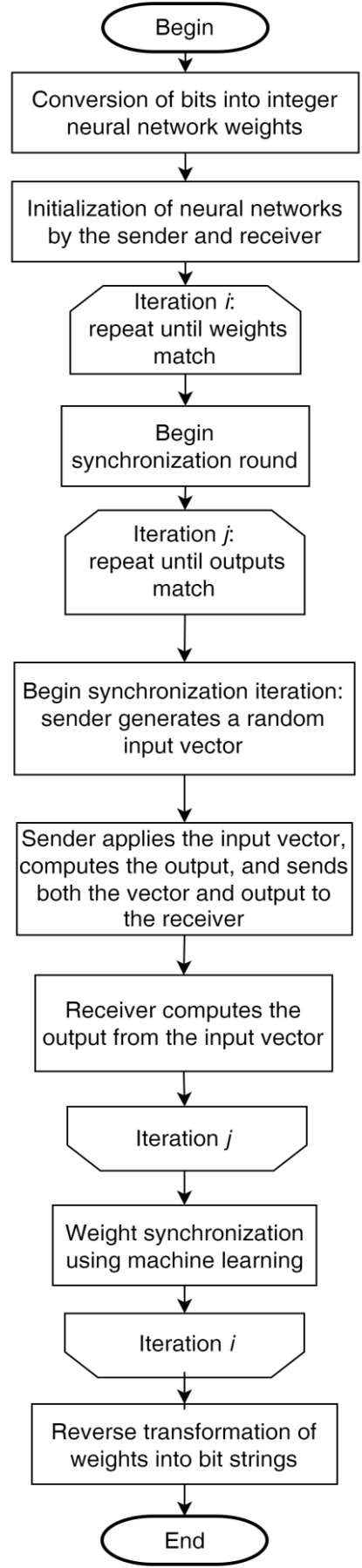

Fig. 2. Block diagram of the generalized key reconciliation algorithm using a neural network.

## III. EXPERIMENTAL INVESTUIGATION

For a quantitative study of the protocol, a software model of the key reconciliation process was created according to the presented algorithm. The study involved a computational experiment consisting of performing a sequence of reconciliation rounds until full agreement was achieved, but with a maximum of 300 iterations. It should be noted that information losses due to sequence comparison between rounds were not taken into account. This is because such comparisons can be performed in different ways depending on the specific task, and their inclusion in the model requires separate consideration. Two experiments were conducted using the implemented model.

### A. Efficiency Metrics

The following metrics were used for a quantitative evaluation of the protocol's performance:

- **Frame Error Rate (FER):** the proportion of bit strings (frames) discarded due to the inability to achieve identical weight coefficients;

- **Amount of Compromised Information (Enthropy Loss):** a measure of the reduction in sequence entropy during an attack by an adversary attempting to guess the sequence based on messages transmitted over the classical (non-quantum) channel.

- **Total number of iterations:** the sum of all iterations performed across all reconciliation rounds;

- **Number of rounds:** the total number of reconciliation rounds required to achieve full agreement between the weight coefficients.

To estimate the amount of entropy loss $Z$, the expression (2) proposed in [7] was used:

$$Z = \log_{2L+1} 2^i, \qquad (2)$$

where $i$ is the number of reconciliation iterations.

### B. Dependence of Reconciliation Efficiency on QBER

The first experiment aimed to establish the relationships between QBER and the number of iterations iii during key reconciliation, as well as between QBER and the amount of leaked information $Z$.

In the experiment, a key of length $B = 600$ bits was created and divided into blocks of $b = 4$ bits. The hidden layer contained $K = 10$ neurons, with $N = 15$ inputs per hidden-layer neuron. The QBER value was varied in the range [0.005, 0.15] with a step of 0.005. For each QBER, the results were averaged over 1000 measurements.

Based on the results of the first experiment, graphs were plotted showing the dependence of the FER on QBER (Fig. 3) and the dependence of leaked information $Z$ on QBER (Fig. 4). Taken together, these dependencies characterize the efficiency of key reconciliation as a function of the quantum channel parameters.

As shown in Fig. 3, there is a relationship between FER and QBER that exhibits a linear trend within the practically significant QBER range of 0.03 to 0.13. At the same time, an increase in QBER leads to a higher number of iterations iii and, consequently, greater information loss. Nevertheless, both FER and leaked information $Z$ remain relatively low when compared, for example, to the CASCADE protocol [4].

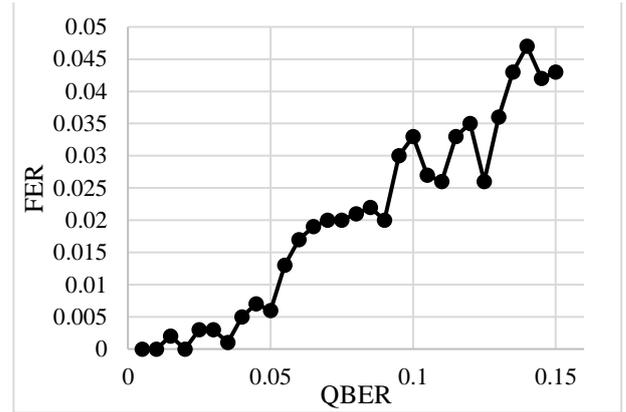

Fig. 3. Dependence of frame error rate (FER) on error rate (QBER).

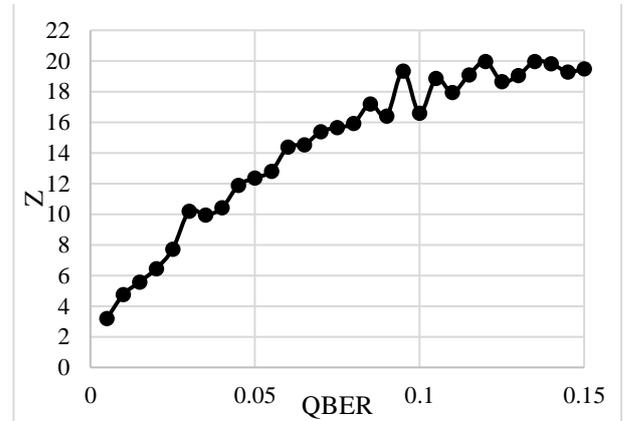

Fig. 4. Dependence of entropy loss $Z$ on error rate (QBER).

The increase in the number of iterations iii with rising QBER is caused, on one hand, by a greater number of non-identical weight coefficients, and on the other hand, by larger differences between these coefficients. Both factors increase the likelihood that individual outputs of the hidden-layer neurons in Alice's and Bob's TPM networks will produce opposite results. As the expected value of differing bit outputs approaches one, the probability of an odd number of differences in the hidden-layer outputs rises, which in turn increases the chance of a mismatch in the overall network output.

### C. Dependence of Reconciliation Efficiency on QBER

The second experiment aimed to determine the relationships between the weight range $L$ and the number of iterations iii during key reconciliation, as well as between $L$ and the amount of leaked information $Z$.

Since $L$ is considered an independent variable, $b$ is calculated according to (1). Consequently, with a fixed network structure ($K = 10$, $N = 15$), the length of the reconciled bit string is determined according to

$$n = b \times K \times N. \qquad (3)$$

The QBER value was kept constant at 0.15.

Based on the results of the second experiment, graphs were plotted showing the dependence of the FER on $L$ (Fig. 5) and the dependence of entropy loss $Z$ on $L$ (Fig. 6). Taken together, these dependencies demonstrate the potential for optimizing the algorithm's efficiency by selecting appropriate parameters.

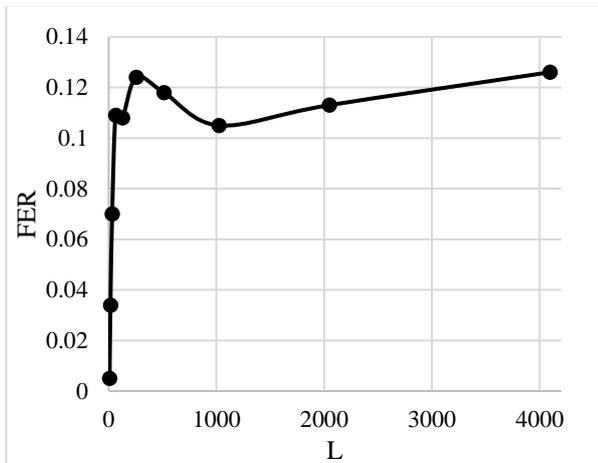

Fig. 5. Dependence of FER on the half-width $L$ of the weight range.

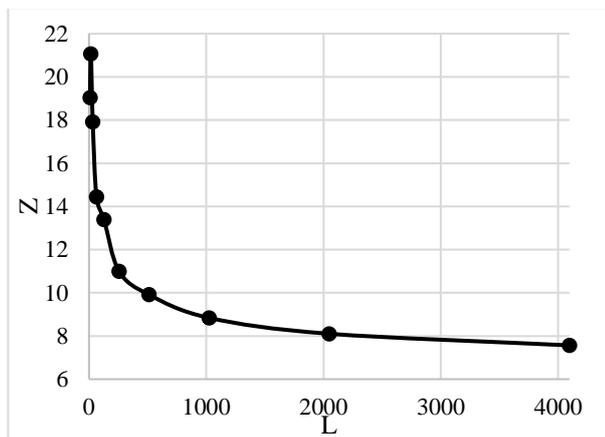

Fig. 6. Dependence of $Z$ on the half-width $L$ of the weight range.

As shown in Fig. 5, increasing $L$ causes a sharp rise in FER within the range [8,128], but the growth rate slows for larger values of $L$. Overall, the observed changes in FER remain moderate, indicating low sensitivity of frame errors to the expansion of the weight range.

The results in Fig. 6 show a hyperbolic relationship. Increasing the weight range initially leads to a sharp decrease in $Z$ up to $L = 256$, where the function exhibits an inflection point. This occurs because, as the set of allowable weight values grows, the number of successful rounds required for synchronization initially increases. However, saturation follows, as the influence of errors in the original bit strings on the hidden-layer neuron outputs weakens, stabilizing the network [10].

Overall, the results of the second experiment indicate that increasing the range of weight coefficients enhances the efficiency of key reconciliation, as implied by (2). When transmitting a single output bit over the classical channel, the amount of leaked information is lower when the entropy of the TPM network's weight values is higher. Therefore, expanding the weight range helps reduce information loss per iteration, partially compensating for the increased number of iterations.

## IV. CONCLUSION

According to the results of the conducted experiments, the TPM-based key reconciliation protocol is functional and demonstrates practical applicability in QKD systems. The data obtained confirm the potential of neural cryptography methods for post-processing key material, as evidenced by the efficiency metrics observed in the experiments, which are comparable to classical parity-check-based methods. At the same time, formal analyses in [10, 11] show that neural cryptography techniques in QKD provide the required level of reliability.

It should be noted that, on one hand, the use of TPM networks allows for parameter optimization through the choice of neural network training algorithm, the range of weight values, and the method of encoding bit strings into weights. On the other hand, the demonstration examples in this work do not account for information losses due to sequence comparison between rounds, which can significantly affect the algorithm's overall efficiency. Therefore, further research in this direction is warranted.

ACKNOWLEDGMENT

The authors express their sincere appreciation for the stimulating discussions, as well as the organizational and intellectual support provided through project FEWM-2023-0015 (TUSUR, Ministry of Science and Higher Education of the Russian Federation).

REFERENCES

[1] A. Zhilyaev et al., *Seti kvantovogo raspredeleniya klyuchey v kiberbezopasnosti*, Goryachaya liniya - Telekom, 2023, 152 p.

[2] C. H. Bennett and G. Brassard, "Quantum cryptography: Public key distribution and coin tossing," *Theoretical Computer Science*, vol. 560, pp. 7–11, 2014.

[3] G. Brassard and L. Salvail, "Secret-Key Reconciliation by Public Discussion," in *Advances in Cryptology – EUROCRYPT '93*, T. Helleseth, Ed. Berlin, Heidelberg: Springer Berlin Heidelberg, 1994, vol. 765, pp. 410–423.

[4] E. V. Antropov et al., "Obzor protokolov ispravleniya oshibok Cascade i AYHI v sistemakh kvantovogo raspredeleniya klyuchey," *Vysokoproizvoditel'nye vychislitel'nye sistemy i tekhnologii*, vol. 8, no. 2, pp. 42–57, 2024.

[5] W. Kinzel and I. Kanter, "Neural Cryptography," *arXiv*, 2002.

[6] M. Stypiński and M. Niemiec, "Weight Equalization Algorithm for Tree Parity Machines," *arXiv*, 2024.

[7] M. Niemiec, "Error correction in quantum cryptography based on artificial neural networks," *Quantum Inf. Process.*, vol. 18, no. 6, p. 174, 2019.

[8] É. Salguero Dorokhin, W. Fuertes, and E. Lascano, "On the Development of an Optimal Structure of Tree Parity Machine for the Establishment of a Cryptographic Key," *Security and Communication Networks*, vol. 2019, pp. 1–10, 2019.

[9] I. Yurchenkov and T. Lishchenko, "Synchronization of Tree Parity Machines," *UniTech*, vol. 122, no. 5, 2024.

[10] A. Ruttor et al., "Genetic attack on neural cryptography," *Phys. Rev. E*, vol. 73, no. 3, p. 036121, 2006.

[11] L. F. Seoane and A. Ruttor, "Successful attack on permutation-parity-machine-based neural cryptography," *Phys. Rev. E*, vol. 85, no. 2, p. 025101, 2012.